\documentclass[preprint,12pt]{elsarticle}

\usepackage{epsfig}
\usepackage{amssymb}

\biboptions{sort&compress}

\begin{document}

\begin{frontmatter}

\title{The Impact of Competing Time Delays in Coupled Stochastic Systems}

\author{D. Hunt\fnref{physics}}
\author{G. Korniss\corref{corrinfo}\fnref{physics}}
\author{B.K. Szymanski\fnref{compsci}}

\address[physics]{Department of Physics, Applied Physics, and Astronomy, \\
Rensselaer Polytechnic Institute, 110 8$^{th}$ Street, Troy, NY
12180--3590, USA}

\address[compsci]{Department of Computer Science \\
Rensselaer Polytechnic Institute, 110 8$^{th}$ Street, Troy, NY 12180--3590, USA}

\cortext[corrinfo]{Corresponding author. korniss@rpi.edu}

\begin{abstract}
We study the impact of competing time delays in coupled stochastic
synchronization and coordination problems. We consider two types of
delays: transmission delays between interacting elements and
processing, cognitive, or execution delays at each element. We
establish the scaling theory for the phase boundary of
synchronization and for the steady-state fluctuations in the
synchronizable regime. Further, we provide the asymptotic behavior
near the boundary of the synchronizable regime. Our results also
imply the potential for optimization and trade-offs in
synchronization problems with time delays.
\end{abstract}


\end{frontmatter}

\section{Introduction}

Coordinating, distributing, and balancing resources in networks is a
complex task as these operations are very sensitive to time delays
\cite{Saber_IEEE2007,Hunt_PRL2010}. To understand and manage the
collective response in these coupled interacting systems, one must
understand the interplay of stochastic effects, network connections,
and time delays. In synchronization, coordination, and consensus
problems in coupled interacting systems
\cite{Arenas_PhysRep2008,GK_Science2003,Nishikawa_PNAS2010,Saber_IEEE2007,Hunt_PRL2010},
individual units attempt to adjust their local state variables
(e.g., pace, load, orientation) in a decentralized fashion. They
interact or communicate only with their local neighbors in the
network, often with explicit or implicit intention to improve global
performance. Applications of the corresponding models range from
physics, biology, computer science to control theory, including
synchronization problems in distributed computing
\cite{GK_Science2003,Guclu_PRE2004,Guclu_FNL2005}, coordination and
control in communication networks
\cite{GK_PRE2007,Hunt_PRL2010,Johari_IEEE2001,Saber_IEEE2004,Saber_IEEE2007,Jadbabaie_IEEE2006,Scutari_IEEE2008},
flocking animals
\cite{Reynolds_ACM1987,Vicsek_PRL1995,Cucker_IEEE2007}, bursting
neurons \cite{Atay_PRL2003,Gassel_FNL2007,Chen_EPL2008,Chen_PRE2009}, and
cooperative control of vehicle formation \cite{Fax_IEEE2004}.

In this Letter, we study the impact of competing, finite non-zero
time delays in stochastic synchronization or coordination problems,
which are present in most real communication, information
\cite{Hunt_PRL2010,Johari_IEEE2001,Saber_IEEE2004,Saber_IEEE2007,Huberman_IEEE1991,Strogatz_PRE2003,Chen_PhysA2004},
and biological systems
\cite{Hutchinson_1948,May_Ecology1973,Ji_FNL2008} including
neurobiological networks
\cite{Gassel_FNL2007,Chen_EPL2008,Chen_PRE2009}. (Throughout this
Letter, we use the terms coordination and synchronization
synonymously.) Delays can be attributed to both non-zero
transmission times between the nodes and to non-zero finite times it
takes to process (possibly including cognitive delays) and execute
the desired action at the nodes. Here, we investigate the importance
and impact of these two types of delays in a simple synchronization
problem in noisy environment with two linearly coupled nodes.

Singularities in critical phenomena and phase transitions \cite{Dorog_RMP2008}, which are
often present in coupled interacting systems consisting of a large
number of nodes $N$, are typically associated with progressively
more eigenvalues of the coupling operator (e.g., the Laplacian)
getting arbitrarily close to zero. Strictly speaking, these
singularities are exhibited only by systems approaching the
thermodynamic limit (where the density of eigenvalues does not
vanish sufficiently fast, or itself becomes singular about zero in
the $N$$\to$$\infty$ limit). For example, in spatially-embedded
physical systems these singularities are typically exhibited by the
relevant response functions and fluctuations in the long-wavelength
limit \cite{GK_Science2003}. In complex networks
\cite{Barab_sci,Watts_Nature1998,BarabREV,MendesREV} these
singularities can be suppressed as a result of sufficient amount of
randomness in the connectivity pattern
\cite{GK_Science2003,Barahona_PRL2002,Kozma_UGA2004,Kozma_PRL2004,Kim_PRL2007}.

In contrast, the instability governed by time delays is associated
with a single mode exceeding a threshold value (in a simple case,
associated with the eigenmode of the network Laplacian with the
largest eigenvalue \cite{Hunt_PRL2010,Saber_IEEE2004}). Therefore,
the underlying instability is present even in the simplest network
with two nodes ($N$=$2$). Here we focus on a two-node network, which
qualitatively captures the generic features of the coordination
behavior when the delays are present due to both transmission
between nodes and processing/execution at each node. For simplicity,
we will refer to this instability as ``critical", even though this
singularity does not require infinitely many degrees of freedom. In
networks, consisting of a large number of nodes, the effect of time
delays is not qualitatively different, but can be ``amplified" by
heterogeneous (or scale-free) \cite{Barab_sci,BarabREV,{MendesREV}}
connectivity patterns: in the case of uniform time delays
\cite{Hunt_PRL2010,Saber_IEEE2007,Saber_IEEE2004}, the effective
coupling of the most relevant singular mode is the largest
eigenvalue of the coupling operator, which itself can diverge with
the system size
\cite{Arenas_PhysRep2008,Fiedler_1973,Anderson_1985,Mohar_1991},
severely limiting synchronizability and coordination.

\section{A Stochastic Model with Local and Transmission Delays}

Differential equations with delays \cite{Bellman_1963} describing
complex systems have a long history, originally motivated by the
emergence of business and economics cycles
\cite{Kalecki_1935,Frisch_1935,Hayes_1950}, and also naturally
appearing in the context of stability of ecological systems, in
models in population dynamics, and in game theory
\cite{Hutchinson_1948,May_Ecology1973,MacDonald_1978,Ruan_2006,Yi_JTB1997,Miekisz_2007}.
There have been recent works combining stochastic differential
equations with delays
\cite{Kuchler_MCS2000,Amann_PhysA2007,Ibanez_FNL2008} with
applications ranging from population dynamics, epidemiology, and
immunology to cell kinetics and finance
\cite{Bocharov_JCAM2000,Tian_JCAM2007}.

Here we consider a model for local coordination where time delays
are attributed to two separate origins: one is the transmission
between the two nodes, the other is processing the information and
executing the action at each node, denoted by $\tau_{\rm tr}$ and
$\tau_{\rm o}$, respectively. We investigate the simplest stochastic
model where the coordination or synchronization attempt between the
two nodes, in terms of the relevant state variables $h_i$, is
captured by linear relaxation
\begin{eqnarray}
\partial_t h_1(t) & = & -\lambda[h_1(t-\tau_{\rm o}) - h_2(t-\tau_{\rm o}-\tau_{\rm tr})] + \eta_1(t) \nonumber \\
\partial_t h_2(t) & = & -\lambda[h_2(t-\tau_{\rm o}) - h_1(t-\tau_{\rm o}-\tau_{\rm tr})] + \eta_2(t) \;.
\label{delay_h_eq}
\end{eqnarray}
Here, $\eta_i$ is delta correlated noise with $\langle\eta_i\rangle
= 0$ and $\langle\eta_i(t)\eta_j(t')\rangle = 2D\delta_{ij}\delta(t
- t')$ with noise intensity $D$, $i,j=1,2$. $\lambda>0$ is the
coupling strength between the two nodes. For initial conditions, we
use $h_i(t)$$\equiv$$0$ for $t$$\leq$$0$.

To simplify notation we introduce $\tau\equiv\tau_o+\tau_{\rm tr}$
and $\gamma\equiv\tau_{\rm o}/(\tau_{\rm o}+\tau_{\rm tr})=\tau_{\rm
o}/\tau$ ($0\leq\gamma\leq1$). Further, since we are interested in
the synchronization (or coordination) between the two nodes, we
focus on the relative difference $u(t)=h_2(t)-h_1(t)$ which is
governed by
\begin{equation}
\partial_t u(t) =  -\lambda u(t-\gamma\tau) - \lambda u(t-\tau) + \xi(t) \;,
\label{delay_u_eq}
\end{equation}
where $\langle\xi\rangle = 0$ and $\langle\xi(t)\xi(t')\rangle =
4D\delta(t - t')$. The special case $\gamma$$=$$1$ of the above
equation has been investigated in our earlier work
\cite{Hunt_PRL2010}. Here, we study the impact of both types of
delays, corresponding to the general case
$0$$\leq$$\gamma$$\leq$$1$. Our quantity of interest is $\langle
u^2(t)\rangle$, capturing the relative deviation of the relevant
state variables on the two nodes. By definition, the system is
synchronizable if the fluctuations reach a finite steady state,
$\langle u^2(\infty)\rangle<\infty$.
In the absence of time delays ($\tau$$=$$0$) one immediately finds
$\langle u^2(t)\rangle=(D/\lambda)(1-e^{-4\lambda t})$ \cite{Gardiner_1985}, i.e., the
system is synchronizable for any $\lambda>0$. Further, the stronger
the coupling, the better the synchronization: $\langle
u^2(\infty)\rangle=D/\lambda$ is a monotonically decreasing function
of $\lambda$.

\begin{figure}[t]
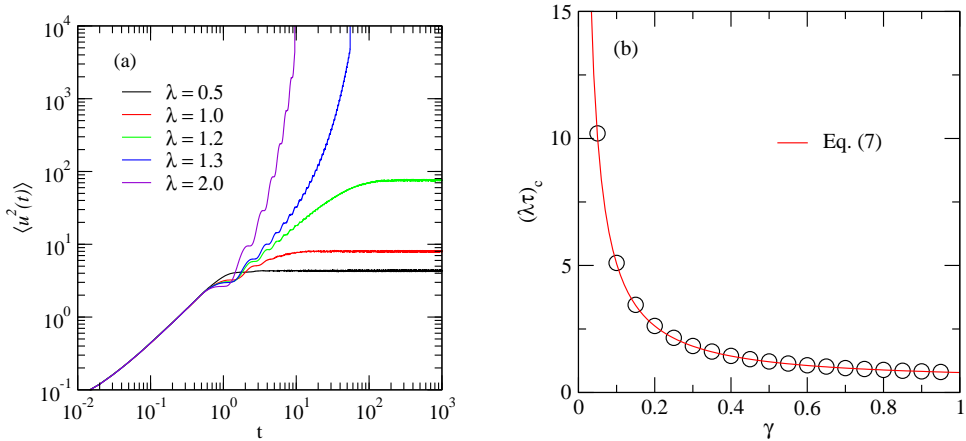

\vspace*{0.80cm}
\centerline{
\psfig{file=fig1a.eps,width=6cm}
\hspace*{0.50cm}
\psfig{file=fig1b.eps,width=6cm}}
\caption{
(a) Time series $\langle u^2(t)\rangle$ for $\tau$$=$$1.00$, $\gamma$$=$$0.50$ for different values
of the coupling constant $\lambda$. Here, and throughout this paper, $D$$=$$1$ and $\Delta t$$=$$0.01$.
(b) Synchronizability threshold in terms of the scaled variable
$\lambda\tau$ vs $\gamma$. Data points were obtained by numerically
integrating Eq.~(\ref{delay_u_eq}) \cite{note_numerical}.  The solid
line represents the exact analytic expression
Eq.~(\ref{phase_diagr}).}
\label{Lc_fig}
\end{figure}

Next we study and analyze the case with time delays. Employing
standard Laplace transform \cite{Hunt_PRL2010,Bambi_JEDC2008}, one can
immediately write the formal solution for Eq.~(\ref{delay_u_eq})
\begin{equation}
u(t) = \int_0^t dt' \xi(t')\sum_\alpha\frac{e^{s_{\alpha}(t - t')}}{h^{'}(s_{\alpha})} \;,
\label{u_solution}
\end{equation}
where $s_{\alpha}$, $\alpha=1,2,\dots$, are the zeros of the characteristic equation
\begin{equation}
g(s) \equiv s + \lambda e^{-\gamma\tau s} + \lambda e^{-\tau s} = 0 \;
\label{char_eq}
\end{equation}
on the complex plane. Then for the noise-averaged fluctuations one finds
\begin{eqnarray}
\langle u^2(t)\rangle & = & \sum_{\alpha,\beta} \frac{-4D(1-e^{(s_\alpha + s_\beta) t})}{g^{'}(s_\alpha)g^{'}(s_\beta)(s_\alpha + s_\beta)} \nonumber \\
                      & = & \sum_{\alpha,\beta} \frac{-4D(1-e^{(s_\alpha + s_\beta) t})}{(1 - \gamma\lambda\tau e^{-\gamma\tau s_{\alpha}} - \lambda\tau e^{-\tau s_{\alpha}})(1 - \gamma\lambda\tau e^{-\gamma\tau s_{\beta}} - \lambda\tau e^{-\tau s_{\beta}})  (s_\alpha + s_\beta)} \nonumber \\
                      & = & \sum_{\alpha,\beta} \frac{-4D\tau(1-e^{(z_\alpha + z_\beta) t/\tau})}{(1 - \gamma\Lambda e^{-\gamma z_{\alpha}} - \Lambda e^{-z_{\alpha}})(1 - \gamma\Lambda e^{-\gamma z_{\beta}} - \Lambda e^{-z_{\beta}})  (z_\alpha + z_\beta)} \;,
\label{u2_equation}
\end{eqnarray}
where in the last expression of the above equation we introduced the
scaled variables $z_{\alpha}\equiv\tau s_{\alpha}$ and
$\Lambda\equiv\lambda\tau$. From Eq.~(\ref{char_eq}) and from the definition of these scaled
variables it is evident that $z_{\alpha}$ are the solutions of the
scaled characteristic equation
\begin{equation}
z + \Lambda e^{-\gamma z} + \Lambda e^{z} = 0 \;,
\label{scaled_char_eq}
\end{equation}
and consequently, the solutions depend only on $\Lambda$, i.e.,
$z_{\alpha}=z_{\alpha}(\Lambda)$. From the structure of the above
characteristic equation it follows that if $z$ is a solution of
Eq.~(\ref{scaled_char_eq}) so is its complex conjugate $z^*$. From
Eq.~(\ref{u2_equation}) it is clear that synchronization can only be
achieved if ${\rm Re}(z_{\alpha})$$<$$0$ for {\em all} $\alpha$. To
identify the boundary of the region of synchronizability, one has to
find the solution(s) with a vanishing real part, i.e., $z=x+iy$ with
$x$$=$$0$ \cite{Frisch_1935,Hayes_1950,May_Ecology1973,Ruan_2006}.
Elementary analysis yields $y^{\pm}_{\rm c}$$=$$\pm\pi/(1+\gamma)$
and
\begin{equation}
(\lambda\tau)_{\rm c} = \Lambda_{\rm c}(\gamma) =
\frac{\pi}{2(1+\gamma)}\frac{1}{\cos(\frac{\pi}{2}\frac{1-\gamma}{1+\gamma})} \;.
\label{phase_diagr}
\end{equation}
Thus, for a fixed $\gamma$, the system is synchronizable if $0 <
\lambda\tau < \Lambda_{\rm c}(\gamma)$. Results obtained by
numerically integrating Eq.~(\ref{delay_u_eq}) \cite{note_numerical} together with the
analytic expression Eq.~(\ref{phase_diagr}) are shown in
Fig.~\ref{Lc_fig}.
We will discuss the phase diagram and some limiting cases in terms
of the original variables, the local delay $\tau_{\rm o}$ and the
transmission delay $\tau_{\rm tr}$, in the final section.

\section{Scaling and Asymptotics in the Steady State}

\begin{figure}[t]
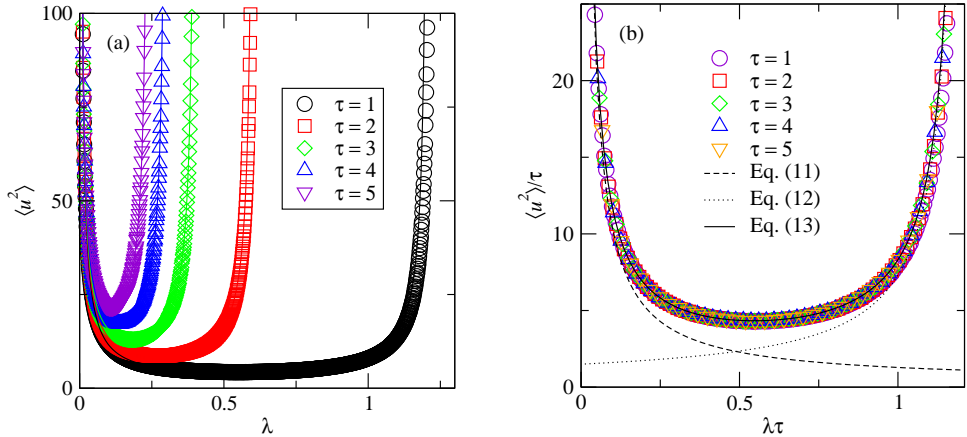

\vspace*{0.80cm}
\centerline{
\psfig{file=fig2a.eps,width=6cm}
\hspace*{0.50cm}
\psfig{file=fig2b.eps,width=6cm}}
\caption{(a) Steady-state fluctuations as a function of the coupling strength $\lambda$
for the various delays for $\gamma$$=$$0.5$.
Data points are obtained by numerically integrating Eq.~(\ref{delay_u_eq}).
(b) Same data as in (a) scaled according to Eq.~(\ref{u2_scaled}), $\langle u^2(\infty)\rangle/\tau$ vs $\lambda\tau$.
The dashed lines represent the asymptotic behaviors of the scaling function near the two endpoints of the synchronizable
regime \cite{note_numerical}, Eqs.~(\ref{f_Lambda0}) and (\ref{f_Lambdac}), respectively,
while the solid line (running precisely through the data points) represents the full approximate scaling function $f(\lambda\tau)$, Eq.~(\ref{f_Lambda_full}).}
\label{u2_fig}
\end{figure}

\begin{figure}[t]
\vspace*{1.00cm}
\centerline{\psfig{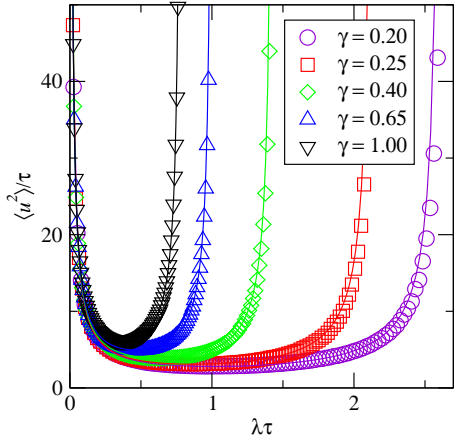}}
\vspace*{8pt}
\caption{Scaled steady-state fluctuations $\langle
u^2(\infty)\rangle/\tau$ vs $\lambda\tau$ for various $\gamma$
values. Data points are obtained by numerically integrating
Eq.~(\ref{delay_u_eq}). Solid lines represent the full approximate
scaling function $f(\lambda\tau)$ for each $\gamma$, Eq.~(\ref{f_Lambda_full}).}
\label{u2_g_fig}
\end{figure}

Now we turn to analyzing the steady-state fluctuations, in
particular, their scaling behavior in the synchronizable regime,
$0<\Lambda<\Lambda_{\rm c}(\gamma)$. Here, the fluctuations remain finite,
and in the steady state ($t$$\to$$\infty$) from
Eq.~(\ref{u2_equation}) one obtains
\begin{equation}
\langle u^2(\infty)\rangle = D\tau f(\gamma,\Lambda) \;,
\label{u2_ss}
\end{equation}
where
\begin{equation}
f(\gamma,\Lambda)= \sum_{\alpha,\beta} \frac{-4}{(1 - \gamma\Lambda e^{-\gamma z_{\alpha}} - \Lambda e^{-z_{\alpha}})
(1 - \gamma\Lambda e^{-\gamma z_{\beta}} - \Lambda e^{-z_{\beta}})  (z_\alpha + z_\beta)}
\label{f_Lambda}
\end{equation}
is the scaling function for the steady-state fluctuations. [Recall
that $z_{\alpha}=z_{\alpha}(\Lambda)$ are the solutions of the
scaled characteristic equation Eq.~(\ref{scaled_char_eq}).] Thus for
a given $\gamma$,
\begin{equation}
\frac{\langle u^2(\infty)\rangle}{D\tau} = f(\lambda\tau) \;,
\label{u2_scaled}
\end{equation}
where in our notation, we suppressed the $\gamma$ dependence to
highlight the scaling behavior of the fluctuations, valid for each
$\gamma$ separately. Figure~\ref{u2_fig} shows the steady-state
fluctuations before (a) and after (b) scaling, and demonstrates the
data collapse for the scaled variables according to
Eq.~(\ref{u2_scaled}).
The scaling function $f(\Lambda)$ is a non-monotonic function of its
argument, diverging at $\Lambda$$=$$0$ and $\Lambda$$=$$\Lambda_{\rm
c}(\gamma)$, and exhibiting a single minimum between these points
[Fig.~\ref{u2_fig}(b)]. This non-monotonic feature of the scaling
function with a single minimum between $0<\lambda\tau<\Lambda_{\rm
c}(\gamma)$ is present for all $0<\gamma\leq$$1$
[Fig.~\ref{u2_g_fig}]. Thus, for fixed non-vanishing and finite
delays, there is an optimal value of the coupling constant $\lambda$
for which the steady-state fluctuation attains its minimum value.
For stronger couplings, the overall coordination between the two
nodes weakens, and for $\lambda>\Lambda_{\rm c}(\gamma)/\tau$, it
completely deteriorates.

Next, we briefly discuss the asymptotic behavior of the scaling
function near the boundaries of the synchronizable regime. The
fluctuations of $\langle u^2(\infty)\rangle$ diverge at the end
points of this interval [as at least for one $\alpha$, ${\rm
Re}(z_{\alpha})\to 0$], indicating the breakdown of synchronization.
Near these endpoints, the sum in Eq.~(\ref{f_Lambda}) is dominated by the
term(s) where ${\rm Re}(z_{\alpha})\simeq0$ \cite{Hod_2010}.
These are the solutions which have (negative) real parts with the smallest amplitude.
As we show in Appendix A, to leading order,
\begin{equation}
f(\Lambda) \simeq \frac{1}{\Lambda}
\label{f_Lambda0}
\end{equation}
as $\Lambda$$\to$$0$, and
\begin{equation}
f(\Lambda) \simeq \frac{c_1(\gamma)}{\Lambda_{\rm c}(\gamma)-\Lambda}
\label{f_Lambdac}
\end{equation}
as $\Lambda$$\to$$\Lambda_{\rm c}$ ($\Lambda$$\lesssim$$\Lambda_{\rm c}$)
with $c_1(\gamma)$ given in
Appendix A [Eq.~(\ref{c1})]. From the numerical results
[Fig.~\ref{u2_fig}(b)] it is also apparent that the scaling function
varies slowly between (and away from) the singular points, thus,
$f(\Lambda)$ can be reasonably well approximated \cite{Hod_2010}
throughout the full synchronizable regime $0<\Lambda<\Lambda_{\rm
c}(\gamma)$ by
\begin{equation}
f(\Lambda) \approx \frac{1}{\Lambda} + \frac{c_1(\gamma)}{\Lambda_{\rm c}(\gamma)-\Lambda} + c_2(\gamma),
\label{f_Lambda_full}
\end{equation}
with $c_2(\gamma)$ also given in Appendix A [Eq.~(\ref{c2})].

Figure~\ref{u2_fig}(b) and Fig.~\ref{u2_g_fig} show that the above
approximate scaling function Eq.~(\ref{f_Lambda_full}) (being
asymptotically exact near the singular points) matches the
numerical data very well. In particular, it captures the basic
non-monotonic feature of the results obtained from numerical
integration, exhibiting a single minimum
\begin{equation}
\Lambda_{\min}(\gamma)=  \frac{\Lambda_{\rm c}(\gamma)}{1+\sqrt{c_{1}(\gamma)}}
\label{Lambda_min}
\end{equation}
in the $0<\Lambda<\Lambda_{\rm c}(\gamma)$ interval.

As can also  be seen in Fig.~\ref{u2_g_fig}, the theoretical
asymptotic behavior, captured by the approximate scaling function
Eq.~(\ref{f_Lambda_full}) becomes less accurate for small $\gamma$
near $\Lambda_{\rm c}(\gamma)$. Other than lacking higher-order
corrections to the asymptotic expressions, this is due in part to
the time discretization in the numerical integration
\cite{note_numerical}. For sufficiently small $\gamma$ values, the
condition $\Delta t<<\gamma\tau$ will not hold, and deviations
between the results of the time-discretized numerical scheme and
those of the continuous system Eq.~(\ref{delay_u_eq}) will become
more significant and noticeable.

\section{Discussion and Summary}

\begin{figure}[t]
\vspace*{1.00cm}
\centerline{\psfig{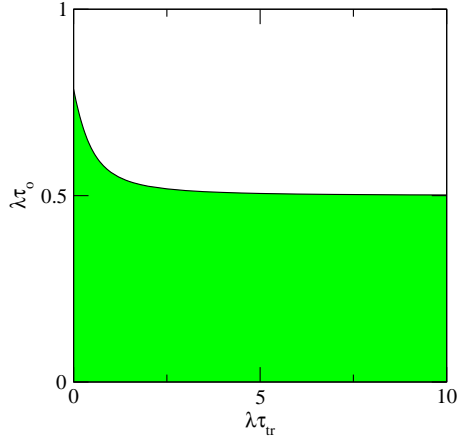}}
\vspace*{8pt}
\caption{Synchronizability phase diagram on the $\lambda\tau_{\rm tr}$-$\lambda\tau_{\rm o}$ plane [Eq.~(\ref{phase_diagr_tau})].
The shaded area indicates the synchronizable regime. The boundary of this region approaches the horizontal line
$\lambda\tau_{\rm o}$$=$$1/2$ in the limit of $\lambda\tau_{\rm tr}$$\to$$\infty$.
Further, $\lambda\tau_{\rm o}$$=$$\pi/4$ when $\lambda\tau_{\rm tr}$$=$$0$.}
\label{phase_diagram_fig}
\end{figure}

Having established the scaling theory for the phase boundary
[Eq.~(\ref{phase_diagr})] and for the fluctuations
[Eq.~(\ref{f_Lambda_full})], it is insightful to express our main
findings explicitly in terms of the two types of delays appearing in
the original formulation of the problem, Eq.~(\ref{delay_h_eq}).
From Eq.~(\ref{phase_diagr}), for the boundary of the synchronizable
regime one immediately finds
\begin{equation}
\lambda(2\tau_{\rm o} + \tau_{\rm tr}) =
\frac{\pi}{2}\frac{1}{\cos(\frac{\pi}{2}\frac{\tau_{\rm tr}}{2\tau_{\rm o}+\tau_{\rm tr}})} \;.
\label{phase_diagr_tau}
\end{equation}
While explicitly expressing the critical line $\tau_{\rm o}$ vs
$\tau_{\rm tr}$ is prohibitive due to the implicit nature of
Eq.~(\ref{phase_diagr_tau}), one can produce a plot for it
numerically without any difficulties [Fig.~\ref{phase_diagram_fig}].

We can gain further insight into the different impact of the two
types of delays by considering two limiting cases. First, consider
the case when $\tau_{\rm o}/\tau_{\rm tr}<<1$, i.e., when the
transmission delays are much larger than the local processing,
cognitive, or execution delays. This is equivalent to the
$\gamma$$<<$$1$ limit in our scaling expressions. From
Eq.~(\ref{phase_diagr}) one finds $\Lambda_{\rm
c}$$\simeq$$1/2\gamma$ or $(\lambda\tau_{\rm o})_{\rm c}$$=$$1/2$.
Thus, there is no singularity in the fluctuations for any finite
$\tau_{\rm tr}$ provided that $\lambda\tau_{\rm o}<1/2$. Further,
from Eq.~(\ref{f_Lambda_full}) (with the coefficients given in
Appendix A) for the steady-state fluctuations in the same limit we
find
\begin{equation}
\langle u^2(\infty)\rangle \simeq \frac{D}{\lambda} +
\left\{ \frac{4}{\pi^2}\frac{1}{1/2-\lambda\tau_{\rm o}} + 1-\frac{8}{\pi^2} \right\}\tau_{\rm tr} \;.
\label{u2_limit1}
\end{equation}

In the other limiting case, $\tau_{\rm tr}/\tau_{\rm o}<<1$, i.e.,
the transmission delays are much smaller than the local processing
delays. This is equivalent to the $\gamma$$\to$$1$ limit in our
scaling expressions. In this limit Eq.~(\ref{phase_diagr}) reduces
to $\Lambda_{\rm c}$$\simeq$$\pi/4$ or $(\lambda\tau_{\rm o})_{\rm
c}=\pi/4$. The steady-state fluctuations approach
\begin{equation}
\langle u^2(\infty)\rangle \simeq \frac{D}{\lambda} +
\left\{ \frac{4}{\pi}\frac{1}{\pi/4-\lambda\tau_{\rm o}} + 2 - \frac{16}{\pi^2} \right\}\tau_{\rm o} \;,
\label{u2_limit2}
\end{equation}
provided that $\lambda\tau_{\rm o}<\pi/4$.

Figure~\ref{phase_diagram_fig} and  Eqs.~(\ref{u2_limit1}) and
(\ref{u2_limit2}) highlight the subtle differences between the
impacts of the two types of delays. We may regard the local delays
$\tau_{\rm o}$ as the dominant ones, in that as long as
$\lambda\tau_{\rm o}<1/2$, there are no singularities for any finite
$\tau_{\rm tr}$, and $\langle u^2(\infty)\rangle$ increases
linearly with $\tau_{\rm tr}$ as $\tau_{\rm tr}$$\to$$\infty$
[Eq.~(\ref{u2_limit1})]. On the other hand, for every $\tau_{\rm
tr}$, there is a sufficiently large $\tau_{\rm o}$ such that the
fluctuations become singular. In particular, when the transmission
delays are much smaller than the local processing delays, the
fluctuations diverge as $\lambda \tau_{\rm o}$$\to$$\pi/4$
[Eq.~(\ref{u2_limit2})].

Inside the synchronizable regime, for fixed $\tau_{\rm o}$ and
$\tau_{\rm tr}$ (with the exception of $\tau_{\rm o}$$=$$0$
\cite{note_min_limit}), the steady-state fluctuations $\langle
u^2(\infty)\rangle$ always exhibit a single local minimum as a
function of the coupling constant $\lambda$
[Eq.~(\ref{Lambda_min})]. This feature naturally presents scenarios
for optimization and trade-offs. In particular, in real systems, the
effective coupling constant between two interacting nodes can be
controlled by the frequency of pairwise communications. This implies
that too much communication can cause more harm than good, and
further, there is an optimal rate of communications between the
nodes which minimizes the size of the fluctuations. Also, as was
already shown for the special $\gamma$$=$$1$ case
\cite{Hunt_PRL2010}, in large network-coupled systems, decreasing
connectivity may be beneficial if the system is too close or beyond
the synchronization boundary.

In this Letter we only considered linear couplings between two nodes
in the presence of noise and competing time delays. We established
the phase diagram and the scaling theory for synchronizability, and
provided the asymptotic behavior for the relevant scaling functions.
Nonlinearities are undoubtedly important in real systems, and will
likely further increase the complexity of the already rich phase
diagram, such as recurring and alternating patterns of
synchronizable and unsynchronizable regions
\cite{Chen_EPL2008,Chen_PRE2009,Strogatz_PRE2003}.

\section*{Acknowledgements}
We thank S. Hod for bringing our attention to his preprint and
subsequently published work on the asymptotic method for the scaling
function \cite{Hod_2010}.
This work was supported in part by DTRA Award No. HDTRA1-09-1-0049,
by the Army Research Laboratory under Cooperative Agreement
Number W911NF-09-2-0053, and by the Office of Naval Research Grant No. N00014-09-1-0607.
The views and conclusions contained in this document are those of the authors and should not be interpreted as
representing the official policies, either expressed or implied, of
the Army Research Laboratory or the U.S. Government.

\appendix

\section{Asymptotic Behavior of the Scaling Function Near the Synchronization Thresholds}

Here, we employ the method in Ref.~\cite{Hod_2010} to calculate the
dominant contributions in Eq.~(\ref{f_Lambda}) near the boundaries
of the synchronizable regime. We assume that solutions of the
characteristic equation
\begin{equation}
z +\Lambda e^{-\gamma z} + \Lambda e^{-z} = 0
\label{char_app}
\end{equation}
change continuously with the parameter $\Lambda$. Thus, if $z$$=$$z_o$
is a solution for $\Lambda$$=$$\Lambda_o$, then for a small change in
the parameter, $\Lambda = \Lambda_o+\delta\Lambda$, the
corresponding solution can be written as $z=z_o+\delta z$.
Substituting this into the characteristic equation, to lowest order
we find
\begin{equation}
\delta z \simeq - \frac{e^{-\gamma z_o} + e^{-z_o}}{1-\gamma\Lambda_o e^{-\gamma z_o} - \Lambda_o e^{-z_o}} \delta\Lambda + {\cal O}\left((\delta\Lambda)^2\right) \;.
\end{equation}

For $\Lambda$$=$$0$, there is a single solution with vanishing real
part, $z$$=$$0$, thus for small $\Lambda$
\begin{equation}
z(\Lambda) \simeq -2\Lambda + {\cal O}(\Lambda^2)\;.
\end{equation}
The dominant contribution for the scaling function as
$\Lambda$$\to$$0$ comes from the corresponding term in
Eq.~(\ref{f_Lambda}), to leading order yielding
\begin{equation}
f(\Lambda) \simeq \frac{-4}{2(-2\Lambda)}  = \frac{1}{\Lambda} \;.
\end{equation}

For $\Lambda$$=$$\Lambda_{\rm c}(\gamma)$ [Eq.~(\ref{phase_diagr})],
there is a pair of solutions (complex conjugates) with vanishing real parts
$z=\pm i\frac{\pi}{1+\gamma}$. When $\Lambda$ is in the
vicinity of $\Lambda_{\rm c}$ ($\Lambda\simeq\Lambda_{\rm
c}+\delta\Lambda$), to lowest order, these solutions behave as
\begin{equation}
z_{\pm}(\Lambda) \simeq \pm iy_{\rm c} - \frac{e^{\mp i\gamma y_{\rm c}}+e^{\mp i y_{\rm c}}}{1-\gamma \Lambda_{\rm c}e^{\mp i \gamma y_{\rm c}}-\Lambda_{\rm c}e^{\mp i y_{\rm c}}} \delta\Lambda \;,
\end{equation}
where $y_{\rm c}=\frac{\pi}{1+\gamma}$. The dominant contributions
for the scaling function as $\Lambda$$\to$$\Lambda_{\rm c}$ then
come from the two terms in Eq.~(\ref{f_Lambda}) when
$(\alpha=\pm,\beta=\mp)$, yielding
\begin{eqnarray}
f(\Lambda) & \simeq & \frac{-8}{(1-\gamma\Lambda_{\rm c}e^{-i\gamma y_{\rm c}}-\Lambda_{\rm c}e^{-iy_{\rm c}} )( 1-\gamma\Lambda_{\rm c}e^{i\gamma y_{\rm c}}-\Lambda_{\rm c}e^{iy_{\rm c}})(z_{+}+z_{-}) } \nonumber \\
           & = & \frac{c_{1}({\gamma})}{\Lambda_{\rm c}(\gamma)-\Lambda} \;,
\end{eqnarray}
where
\begin{equation}
c_{1}(\gamma) =
\frac{4}{(1+\gamma)\Lambda_{\rm c} + (1+\gamma)\Lambda_{\rm c}\cos(\pi\frac{1-\gamma}{1+\gamma}) -\cos(\frac{\pi}{1+\gamma}) - \cos(\frac{\gamma\pi}{1+\gamma}) } \;.
\label{c1}
\end{equation}

Finally, we obtain the approximate scaling function for the full
$0$$<$$\Lambda$$<$$\Lambda_{\rm c}(\gamma)$ interval, using some
heuristics following Ref.~\cite{Hod_2010}. As observed from our
numerical results [Fig.~\ref{u2_fig}], the scaling function varies
slowly between (and away from) the two singular points. Then, it can
be approximated by
\begin{equation}
f(\Lambda) \simeq \frac{1}{\Lambda} + \frac{c_1({\gamma})}{\Lambda_{\rm c}(\gamma)-\Lambda}  + c_2(\gamma) \;.
\label{f_Lambda_app}
\end{equation}
In principle, the constant $c_{2}(\gamma)$ could be
determined by matching the minimum value of the scaling function.
Since it is not known analytically, instead we resort to the
heuristics of Ref.\cite{Hod_2010} where the constant $c_{2}(\gamma)$
is determined in such a way that it matches next-to-leading order
corrections of the asymptotic behavior, e.g., near $\Lambda$$=$$0$.
To that end, we find the next-to-lowest order corrections to
the solution of Eq.~(\ref{char_app}) in the vicinity of $\Lambda$$=$$0$,
\begin{equation}
z(\Lambda) \simeq -2\Lambda - 2(1+\gamma)\Lambda^2 + {\cal O}(\Lambda^3)\;.
\end{equation}
Keeping the relevant orders in the dominant term in Eq.~(\ref{f_Lambda}), we obtain
\begin{eqnarray}
f(\Lambda) & \simeq & \frac{-4}{(1-\gamma\Lambda-\Lambda)^2\; 2(-2\Lambda - 2(1+\gamma)\Lambda^2)} \nonumber \\
           & \simeq & \frac{1}{[1-(1+\gamma)\Lambda]^2 \Lambda(1 +(1+\gamma)\Lambda)} \nonumber \\
           & \simeq & \frac{1}{\Lambda}+(1+\gamma) \;.
\end{eqnarray}
In order to match this next-to-leading order correction as
$\Lambda$$\to$$0$ with the proposed approximate scaling function
Eq.~(\ref{f_Lambda_app}), one must have
\begin{equation}
c_2(\gamma)  = 1+\gamma - \frac{c_1({\gamma})}{\Lambda_{\rm c}(\gamma)} \;.
\label{c2}
\end{equation}


\begin{thebibliography}{0}
\bibitem{note_numerical}
The time discretization of Eq.~(\ref{delay_u_eq}) naturally has its
own effects on the stability through the numerical scheme. Choosing
$\Delta$$t$$<<$$\gamma\tau$ and  $\Delta$$t$$<<$$1/\lambda$ will
yield only small corrections to the behavior of the underlying
continuous-time system.
\bibitem{note_min_limit}
In the limit of $\tau_{\rm o}<<\tau_{\rm tr}$,
$\lambda_{\min}\simeq\frac{\pi}{4\sqrt{\tau_{\rm o}\tau_{\rm tr}}}$, as can be
obtained from Eq.~(\ref{Lambda_min}).
\end{thebibliography}
\end{document}